# Vector Spin Seebeck Effect and Spin Swapping Effect in Antiferromagnetic Insulators with Non-Collinear Spin Structure

Jinsong Xu[1*], Weiwei Lin[2], Jiaming He[3], J.-S. Zhou[3], Danru Qu[4], Ssu-Yen Huang[5], and C.L. Chien[1,5*]

[1] *Department of Physics and Astronomy, Johns Hopkins University, Baltimore, Maryland 21218, USA*

[2] *Key Laboratory of Quantum Materials and Devices of Ministry of Education, School of Physics, Southeast University, Nanjing 211189, China*

[3] *Department of Mechanical Engineering, University of Texas at Austin, Austin, Texas 78712, USA*

[4] *Center for Condensed Matter Sciences, National Taiwan University, Taipei 10617, Taiwan*

[5] *Department of Physics, National Taiwan University, Taipei 10617, Taiwan*

*Corresponding author. Email: jxu94@jhu.edu (J.X.); clchien@jhu.edu (C.L.C.)

**Abstract**

**Antiferromagnets (AFs) are prospectives for next-generation high-density and high-speed spintronic applications due to their negligible stray field and ultrafast spin dynamics, notwithstanding the challenges in detecting and manipulating AF order with no magnetization ($M = 0$). Among the AFs, non-collinear AFs are of particular interest because of their unique properties arising from the non-collinear spin structure and the small magnetization $M$. In this work, we describe the recently observed vector spin Seebeck effect (SSE) in non-collinear LuFeO$_3$, where the magneto-thermovoltage under an in-plane temperature gradient, not previously observed, is consistent with the predicted spin swapping effect. Our results shed light on the importance of the non-collinear spin structure in the emerging spin phenomena in non-collinear AFs and offer a new class of materials for AF spintronics and spin caloritronics.**

## I. INTRODUCTION

Ferromagnets (FMs) with collinear moments and a large magnetization $M$ have dominated spintronics and spin caloritronics [1,2]. The two directions of ± $M$ of FM have been the basis for magnetic storage. A variety of magneto-transport effects (anomalous Hall effect, anisotropic magnetoresistance, tunnelling magnetoresistance, spin Hall magnetoresistance, among others) in addition to magnetometry can detect the large magnetization $M$, which can be manipulated by spin-transfer torque, spin-orbit torque and magnetic field. However, the large $M$ is also susceptible to external magnetic field and stray field interference. In contrast, antiferromagnetic (AF) materials with zero or nearly zero magnetization are immune to the stray field issues. Some AFs also harbor high frequency excitations beneficial for high-speed devices. However, it is challenging to detect and manipulate AFs with $M = 0$ [3–9]. There have been reports of electrical switching and detection of the AF Néel vector in collinear AF metals and insulators [10–13], where the origin of the electrical signals remains controversial, including some nonmagnetic thermal contributions and electromigration [14–19].

Spin Seebeck effect (SSE) and spin pumping, the two well-established spin injection methods in FMs, are not feasible in collinear AF insulators with $M = 0$, unless acquiring a small induced $M$ by a large magnetic field or beyond the spin-flop field of some AFs [20–23]. In contrast, non-collinear AF with an inherent small $M$ have revealed new spintronics phenomena. For example, in non-collinear AF semimetals of Kagome AF $Mn_3Sn$, the Berry curvature associated with the non-collinear spin structure enables spin-orbit torque switching as evidenced by the anomalous Hall effect, anomalous Nernst effect, and AF tunnelling magnetoresistance [24–31]. Short-wavelength coherent magnon propagation has been demonstrated in non-collinear AF insulating orthoferrite $DyFeO_3$ [32]. Long-distance spin transport has also been reported in the same class of material $YFeO_3$ [33]. In non-collinear AF insulating orthoferrites of $LaFeO_3$ and $LuFeO_3$ [34,35], vector SSE and spin swapping effect have been discovered, as described in the following. The non-collinear AFs offer a unique platform to explore new spin current phenomena that are not viable in collinear FMs and AFs [34–38], and open new grounds for AF spintronic and spin caloritronic applications without the need for a large magnetic field.

Within the SSE, it has been well established in FM insulators with collinear moments that there is only longitudinal SSE but no transverse SSE, with spin injection administered via a temperature gradient in the out-of-plane and the in-plane direction, respectively. In contrast, vector SSE has been observed in non-collinear AF LuFeO$_3$ and LaFeO$_3$, where spin injection via temperature gradient may be administered in any direction, out-of-plane as well as in-plane. The unusual characteristics of the vector SSE are intimately linked to the non-collinear spin structure leading to a small and spontaneous *M*. The characteristics of the unprecedented magneto-thermovoltage under an in-plane temperature gradient in LuFeO$_3$ and LaFeO$_3$ are consistent with those of the predicted spin swapping effect [39]. The magnitude of the vector SSE in LuFeO$_3$ increases greatly by 10-fold from 300 K and peaks at about 400 K. These results highlight the unique role of a new class of non-collinear materials for AF spintronics and spin caloritronics without the limitations of ferromagnets and antiferromagnets with collinear moments.

## II. EXPERIMENTAL DETAILS

**LuFeO$_3$ crystal growth and device fabrication**: Single crystals of LuFeO$_3$ were grown by the floating zone method with an image furnace (NEC SC-M35HD); the starting dense ceramic rods were first prepared by solid state reaction of Lu$_2$O$_3$ (Alfa Aesar, 99.99%) and Fe$_2$O$_3$ (Alfa Aesar, 99.99%) calcined in air for 12 hours in 1200 °C repeatly with intermediate grinding. The crystal growth has been carried out at a rate of about 8 mm/h in a continuous flow of oxygen. The LuFeO$_3$ single crystals are black in color with shiny surfaces. The resulting single crystals were oriented with Laue back reflection. We cut *c*-axis oriented crystals and polished them with lapping papers from 30 μm to 0.05 μm. For the SSE experiments with the W/LuFeO$_3$ devices, we deposited a 3-nm layer of heavy metal W layer onto the polished LuFeO$_3$ crystal disks by DC magneton sputtering at room temperature in a vacuum chamber with a base pressure of $1.0 \times 10^{-8}$ Torr.

**Magnetization characterization**: We used a vector vibrating sample magnetometer (VSM) with two sets of sensing coils, parallel and perpendicular to the applied magnetic field, to simultaneously measure the magnetizations of LuFeO$_3$ along different directions. We have also measured the magnetization of LuFeO$_3$ single crystals with the magnetic field along its *a*-axis, *b*-axis, and *c*-axis using a Magnetic Property Measurement System (MPMS) from Quantum Design.

**Spin Seebeck effect (SSE) measurements**: We used two SSE setups for longitudinal SSE and transverse SSE measurements, in which the temperature gradient has been applied exclusively in the out-of-plane and in-plane directions, respectively. We thermally anchored the W/LuFeO$_3$ devices to two Cu blocks via silicone thermal pads for better thermal contact, and one of the Cu blocks was heated by a resistive heater. The temperatures of the two Cu blocks ($T_{\text{hot}}$ and $T_{\text{cold}}$), as monitored by two Cernox temperature sensors, give the temperature difference $\Delta T \equiv T_{\text{hot}} - T_{\text{cold}}$, and the sample temperature $T_{\text{avg}} \equiv (T_{\text{hot}} + T_{\text{cold}})/2$. We express the temperature gradient as $\nabla T \equiv \Delta T/L_{\text{T}}$, where $L_{\text{T}}$ is the sample length between the two temperatures with typical values of $\nabla T$ on the order of 5 K/mm. We express the magneto-thermal voltages as $\Delta V/L_{\text{V}}$, where $L_{\text{V}}$ is the separation of the voltage leads. Since the SSE voltage is proportional to the temperature gradient, we display the results as $(\Delta V/L_{\text{V}})/(\Delta T/L_{\text{T}})$, taken into account the dimensions of the samples and thermal gradients.

## III. RESULTS AND DISCUSSION

**Longitudinal spin Seebeck effect in collinear ferromagnetic insulators**

We first illustrate the well-known longitudinal SSE results in FM insulators using the W/YIG devices, where an out-of-plane temperature gradient injects a pure spin current $J_s$ into the W layer. The inverse spin Hall effect (ISHE) in the W layer generates an electric field of

$$\boldsymbol{E}_{\text{ISHE}} = (\theta_{\text{SH}}\rho)\boldsymbol{J}_s \times \boldsymbol{\sigma} \qquad (1)$$

and detected as an ISHE voltage $V$, where $\theta_{\text{SH}}$ and $\rho$ are the spin Hall angle and resistivity of W, respectively. It has been well established that there is only longitudinal SSE but no transverse SSE in collinear FMs and AFs [40–46]. In W/YIG, under an out-of-plane temperature gradient $\nabla_{\mathbf{z}}T$ and an in-plane magnetic field $H_y$, which aligns the spin index $\boldsymbol{\sigma}$ of YIG along the $y$-direction, one detects the ISHE voltage in the $x$-direction but not in the $y$-direction, according to Eq.(1) as shown in Fig. 1a. In FM insulators, such as YIG, a modest magnetic field exceeding the anisotropy field can readily align the magnetic moments to any direction and establish $\boldsymbol{\sigma}$ in that direction.

On the other hand, as shown in Fig. 1b, for the transverse SSE scheme in W/YIG under an in-plane temperature gradient $\nabla_{\mathbf{y}}T$, there is no discernible magneto-thermovoltage $V_{\text{x}}$ for the magnetic field applied along the $z$-direction (0°), the $y$-direction (90°), and in between (45°), indeed along any direction. This is because the W overlayer can detect only pure spin current with an in-plane spin index that has been injected in the out-of-plane direction. One notes that null results of transverse SSE in W/YIG shown in Fig. 1b attest to an exclusive in-plane temperature gradient.

To date, only the longitudinal SSE under an out-of-plane $\nabla T$ has been experimentally observed in collinear FMs and AFs, where an in-plane external magnetic field establishes the spin index $\sigma$, with the ISHE voltage observed in the third direction according to Eq.(1). In this respect, the vectorial relationship of Eq.(1) has been realized only in one direction. On the other hand, very different behavior, a vector SSE, has been observed in *non-collinear* AFs.

**Non-collinear antiferromagnetic spin structure in LuFeO$_3$**

LuFeO$_3$ belongs to the orthoferrite family $R$FeO$_3$ ($R$ = rare earth) with an orthorhombic *Pbnm* (No. 62) crystal structure with lattice parameters $a$ = 0.52 nm, $b$ = 0.55 nm, and $c$ = 0.76 nm. Because of the strong spin-spin exchange interaction, Fe$^{3+}$ moments are aligned antiparallel along the *a*-axis with $T_N$ = 620 K. Due to the Dzyaloshinskii-Moriya interaction (DMI) and/or single ion anisotropy (SIA), a non-collinear spin structure exists in these orthoferrites, exhibiting a weak ferromagnetism [47]. When the rare earth element carries a magnetic moment, the orthoferrites show more complicated and exotic magnetic phases and phenomena, such as spin-reorientation due to the interaction between $R^{3+}$ and Fe$^{3+}$ moments [48–50]. In LuFeO$_3$, Lu$^{3+}$ has no moment with a full 4$f$ shell, whereas Fe$^{3+}$ carries a moment of 5 $\mu_B$, which are largely antiparallel aligned along the *a*-axis but with a small canting angle of about 0.8° towards the *c*-axis. This small canting gives rise to the non-collinear spin structure essential for the vector SSE. The non-collinear spin structures of LuFeO$_3$ are shown in Fig. 2a, where all the moments, largely along the *a*-axis, tilt slightly up for + $M_z$ or slightly down for - $M_z$, depending on the field applied along the *c*-axis. The net moments $M$, which sets the spin index $\sigma$, is along the *c*-axis. Only the magnetic field or component of $H$ along the *c*-axis ($H_z$) induces switching between ± $M$. In subsequent SSE measurements, we applied a magnetic field along the *c*-axis.

To characterize the non-collinear spin structure of LuFeO$_3$, we conducted vector VSM measurements, where the magnetization along two orthogonal directions were measured simultaneously by two sets of sensing coils. In *c*-oriented LuFeO$_3$, the measuring direction *z* is nearly parallel to the *c*-axis of the orthorhombic crystal. The magnetizations $M_x$ and $M_z$ were measured with field $H_z$ (Fig.2c) and $H_x$ (Fig.2d). Likewise, the magnetizations $M_y$ and $M_z$ were measured with field $H_z$ (Fig.2f) and $H_y$ (Fig.2g). With $H_z$, a square hysteresis loop shows $\pm M_z$ with a sharp switching at $\pm 150$ Oe with $M_z \approx 0.05$ $\mu_B$/Fe (Fig. 2c and 2f), which is only about 1% of the full Fe moment. In addition, there is also simultaneous switching in the much smaller $M_x$ and $M_y$, due to the small misalignment between *c*-axis and *z*-direction. Similar behaviors are observed with the applied fields of $H_x$ and $H_y$, but with a much larger switching field, as shown in Fig. 2d and Fig. 2g. Therefore, there are magnetizations in the *c*-oriented $LuFeO_3$ along all three directions with $M_z$ much larger than $M_x$ and $M_y$, essential for the observation of vector SSE.

We measured the temperature dependence of the magnetization of $LuFeO_3$ with a field along the three crystal axes. Fig. 3a shows the results at the low field of $H = 100$ Oe. The magnetization along *c*-axis $M_c$ displays a temperature dependence of $M = A(T_N - T)^\beta$, very similar to those of FMs, that can be well fitted with the critical exponent $\beta = 0.33$ and ordering temperature $T_N = 620$ K. This aspect is similar to those of the FMs, except the value of $M_c$ is two orders of magnitude smaller than the total magnetization if the $Fe^{3+}$ moments were fully aligned. On the other hand, the magnetizations of $M_a$ and $M_b$ along *a* and *b* respectively do not exhibit such a behavior. At a much larger field of $H = 50000$ Oe, the magnetization $M_a$ along *a*-axis exhibits a peak feature (Fig. 3b), a characteristic behavior of antiferromagnetism. Because of the strong AF interaction, the net *M* along the *c*-axis cannot be significantly altered by a magnetic field.

**Vector spin seebeck effect in LuFeO3**

The vector SSE results in *c*-oriented $LuFeO_3$ with field applied along *z*-direction are shown in Fig. 4. Firstly, we observed longitudinal SSE, a magneto-thermovoltage under an out-of-plane temperature gradient $\nabla_z T$ (Fig. 4a and 4b). The longitudinal SSE signal, on the order of ~ 0.5 nV/K at $T_{avg} = 325$ K, exists in the *x*-direction (Fig. 4a) and also in the *y*-direction (Fig. 4b), different from those in W/YIG shown in Fig. 1a. Even more surprising, we have observed a magneto-thermovoltage under an in-plane temperature gradient $\nabla_y T$ (Fig. 4c) and also $\nabla_x T$ (Fig. 4d) with a much larger signal of ~ 4.5 nV/K at $T_{avg} = 320$ K, instead of the null signals in W/YIG as shown Fig. 1b. These are the characteristics of the vector SSE. The spin index $\sigma$ in $LuFeO_3$ is dictated by $M = M_x \mathbf{i} + M_y \mathbf{j} + M_z \mathbf{k}$, where $\mathbf{i}$, $\mathbf{j}$, and $\mathbf{k}$ are the unit vectors along x, y, and z axes, respectively. According to Eq.(1), we have under an out-of-plane $\nabla_z T$,   $(V_x/L_V)/(\Delta T_z/L_T) = aM_y$, and $(V_y/L_V)/(\Delta T_z/L_T) = aM_x$ under an in-plane $\nabla_y T$,   $(V_x/L_V)/(\Delta T_y/L_T) = bM_z$, under an in-plane $\nabla_x T$,   $(V_y/L_V)/(\Delta T_x/L_T) = bM_z$.

where $a$ and $b$ are the common factors for out-of-plane and in-plane $\nabla T$ scheme respectively, and within the same order of magnitude. Magnetometry results reveal $M_z \gg M_x$ and $M_y$ in c-oriented LuFeO$_3$. Consequently, the longitudinal SSE of LuFeO$_3$ under out-of-plane $\nabla T$ that scales with $M_x$ and $M_y$ is much smaller than the magneto-thermovoltage under in-plane $\nabla T$ that scales with $M_z$.

Unlike those in collinear AFs, where a large magnetic field (on the order of several Tesla) is required for SSE, the vector SSE in c-oriented LuFeO$_3$ is readily visible even at zero magnetic field due to the small spontaneous magnetization, which can be reversed by a small switching field of 150 Oe. These non-collinear AFs may provide a promising playground for low-field AF spintronics and spin caloritronics.

**Spin swapping effect**

In c-oriented W/LuFeO$_3$, the small misalignment between the spin index direction along c, and the our-of-plane direction along z, gives rise to the small longitudinal SSE results as shown in Fig. 4a and 4b. Even more fascinating is the presence of much larger magneto-thermovoltage under an in-plane temperature gradient $\nabla_x T$ (Fig. 4c) and $\nabla_y T$ (Fig. 4d), which are completely absent in collinear FMs (e.g., YIG) or AFs (e.g., Cr$_2$O$_3$), and with a magnitude of 4.5 nV/K at $T_{avg} = 320$ K, much greater than those of the longitudinal SSE. The discovery of vector SSE in c-oriented LuFeO$_3$ may be exploited for lateral spintronic phenomena and new device architectures for spintronics and spin caloritronics.

The characteristics of the observed magneto-thermovotlage under an in-plane temperature gradient in c-oriented LuFeO$_3$ is consistent with the spin swapping effect predicted by Lifshits and Dyakonov in 2009 [34,39]. As shown in Fig. 5, a temperature gradient $\nabla_y T$ applied along the y-direction of *c*-oriented LuFeO$_3$ generates a spin/magnon current $J_y$ flowing along the *y*-direction with a spin index $\sigma_z$ along the *z*-direction. The ISHE voltage $V_x$ measured in the W layer detects a spin current $J_z$ flowing along the *z*-direction with a spin index $\sigma_y$ along the *y*-direction. Specifically, the injected $J_y$ with $\sigma_z$ has been converted into $J_z$ with $\sigma_y$ in *c*-oriented LuFeO$_3$ and/or its interface with W. This is consistent with the spin swapping effect, where the directions of the spin index and the spin current interchanged. This interchange of directions between spin current and spin index can arise from magnon-magnon or magnon-phonon scattering in the presence of spin-orbit coupling [51–53]. Experimentally, the spin swapping effect has been realized in non-collinear AF of LuFeO$_3$ and LaFeO$_3$ [34,35], but not in collinear FMs or AFs. Indeed, the spin swapping effect has been proposed in non-magnetic metals with strong spin-orbit coupling [39]. Experimentally, spin currents injected from collinear FMs and AFs do not, and thus far those from non-collinear AFs do exhibit the spin swapping effect.

**Temperature dependence of vector spin Seebeck effect**

In SSE, the temperature gradient injects the spin current from the FM (e.g., YIG) or the non-collinear AF with a net *M* (e.g., LuFeO$_3$) into the adjacent heavy metal (e.g., W) and be detected. The direction of *M* sets the spin index in the specific spin injection. The magnitude of the spin injection at a fixed temperature depends on the relevant component of *M*. However, at different temperatures, the factors *a* and *b* also play an important role, specifically $aM_x$ and $aM_y$ for an out-of-plane temperature gradient and $bM_z$ for an in-plane temperature gradient. The vector SSE results in *c*-oriented LuFeO$_3$ at several temperatures are shown in Fig. 6. Representative curves at three temperatures illustrate the strong temperature dependence of magneto-thermovoltages under out-of-plane temperature gradient (Fig. 6a) and in-plane temperature gradient (Fig. 6b). As shown in Fig. 6c, from 300 K to 400 K, magneto-thermovotlages under out-of-plane and in-plane temperature gradient both increase by more than 10-fold and peaks at 400 K. As shown in Fig. 3a, the values of $M_x$, $M_y$, and $M_z$ actually decrease slightly from 300 K to 400 K. Thus, the sharp increase of the vector SSE, despite the decrease of *M*, comes from the strong temperature dependence of the factors *a* and *b*, due to the strong temperature dependence of magnon properties, magnon-phonon interaction, and spin mixing conductance at the interface [21,22,54–57]. Due to the entropy, the transport of thermal magnons is known to be vanishing towards zero temperature and above the Néel temperature, and thus peaks at a temperature in between. Further studies are required to account for the rapidly increasing vector SSE that peaks at about 400 K.

## IV. CONCLUSION

Previously, only longitudinal SSE under out-of-plane temperature gradient and no transverse SSE under in-plane temperature gradient has been observed in FMs with a large magnetization $M$, whereas no SSE can be realized in AFs with $M = 0$ unless under a large magnetic field. We have observed vector SSE, with both an out-of-plane and an in-plane temperature gradient, in AF insulator of $LuFeO_3$ with a small $M$ along its $c$-axis due to its intrinsic non-collinear spin structure. We show the characteristics of the vector SSE and the close relationship with the various components of $M$ of the non-collinear spin structure. The observed magneto-thermovoltage in $LuFeO_3$ under an in-plane temperature is consistent with the predicted spin swapping effect, where the spin index and the spin current directions are interchanged. The vector SSE in $LuFeO_3$ has a very strong temperature dependence, with magnitude increasing by 10-fold from 300 K to 400 K, while $M$ decreases slightly in the same temperature range. The vector SSE in these non-collinear AFs, readily observable at a zero magnetic field, can be switched by a small magnetic field of about 200 Oe at $T = 300 \sim 450$ K. These non-collinear AFs may enable new device architectures and open new directions for low-field spintronics and spin caloritronics.

## ACKNOWLEDGMENTS

This work was supported by DOE Basic Energy Science Award No. DE-SC0009390, NSF DMREF Award No. 1729555 and 1949701. W. L. acknowledges support from the National Natural Science Foundation of China Grant No. 12074065. D. Qu and S.Y.H. was supported by the Ministry of Science and Technology of Taiwan under Grant No. MOST 110-2123-M-002-010 and No. MOST 110-2112-M-002-047-MY3.

## AUTHOR DECLARATIONS

Conflict of Interest

The authors have no conflicts to disclose.

## DATA AVAILABILITY

The data that support the findings of this study are available from the corresponding author upon reasonable request.

**FIGURE LEGENDS**

FIG. 1 . (a) Longitudinal SSE in W/YIG with blue (red) curve for voltage $V_x$ ($V_y$) measured along $x$-direction ($y$-direction) with magnetic field $H_y$ applied along $y$-direction. (b) Absence of transverse SSE in W/YIG at angles $\beta = 0°$, $45°$, and $90°$ between the applied magnetic field $H$ and $y$-direction.

FIG. 2. (a) Non-collinear spin structure of LuFeO$_3$ with only Fe$^{3+}$ moments shown, where an applied field along $c$-axis can switch the spin orientation. (b) Vector VSM measurements of $M_x$ and $M_z$ in $c$-oriented LuFeO$_3$, with results shown in (c) for magnetic field $H_z$ and (d) for magnetic field $H_x$. (e) Vector VSM measurements of $M_y$ and $M_z$ in $c$-oriented LuFeO$_3$, with results shown in (f) for magnetic field $H_z$ and (g) for magnetic field $H_y$.

FIG. 3. Temperature dependence of magnetization along $a$-axis $M_a$ (red), $b$-axis $M_b$ (blue), and $c$-axis $M_c$ (black) with $H$ applied along corresponding crystal axis at (a) 100 Oe, and (b) 50000 Oe. The orange curve in (a) is the fitting result using $M = A(T_N - T)^\beta$ with $\beta = 0.33$.

FIG. 4. Longitudinal SSE in $c$-oriented W/LuFeO$_3$ under $\nabla_z T$ at $T_{avg}$ = 325 K, (a) $V_x$ measured along $x$-direction (b) $V_y$ measured along $y$-direction. Magneto-thermovoltage in $c$-oriented W/LuFeO$_3$ at $T_{avg}$ = 320 K, (c) $V_x$ under $\nabla_y T$ and (d) $V_y$ under $\nabla_x T$.

FIG. 5. Schematic of spin swapping effect.

FIG. 6. Magneto-thermovotlages in $c$-oriented LuFeO$_3$ at different temperatures, under (a) an out-of-plane and (b) an in-plane temperature gradient. (c) Temperature dependence of magneto-thermovoltages in $c$-oriented LuFeO$_3$ under an out-of-plane (blue) and an in-plane (black) temperature gradient.

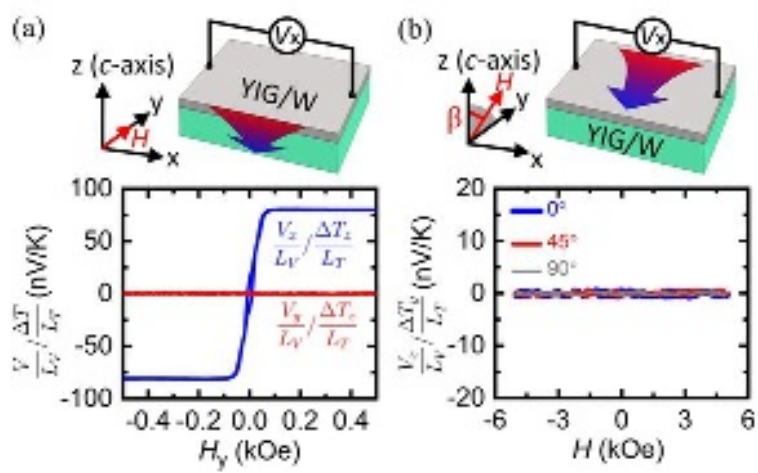

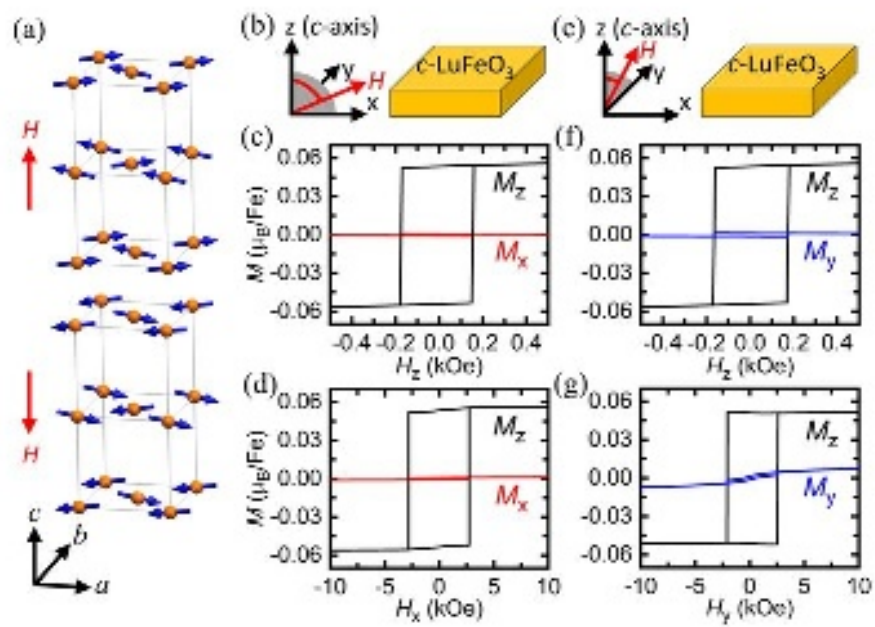

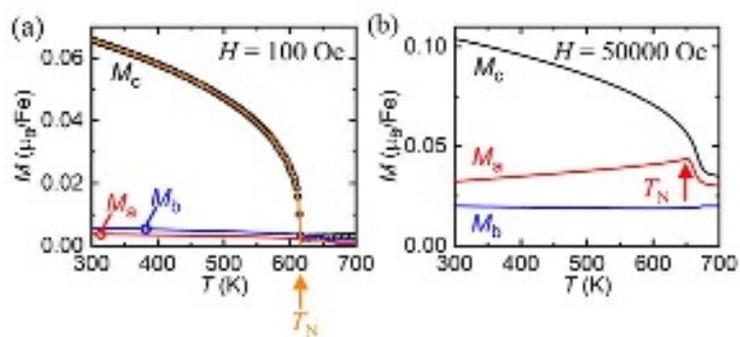

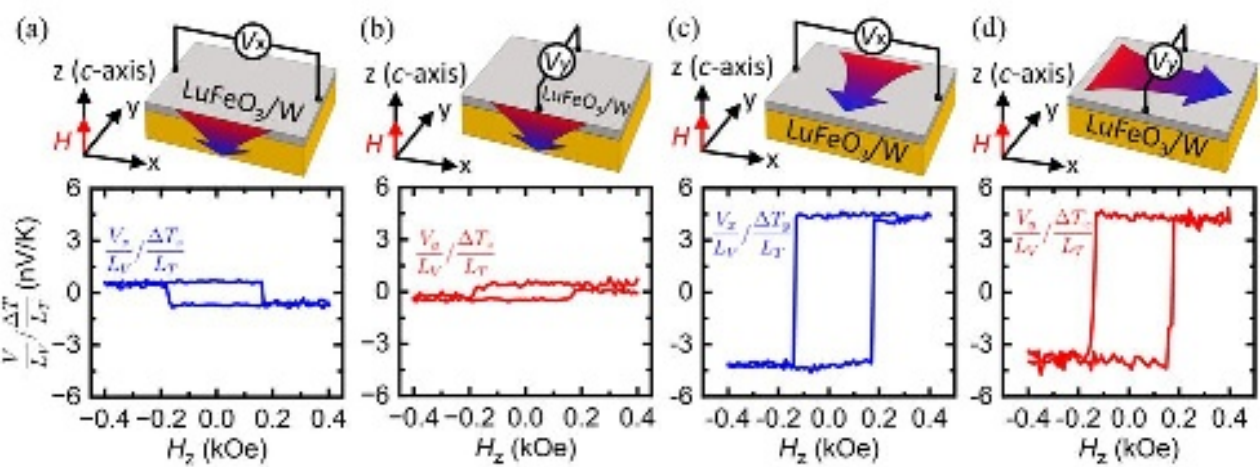

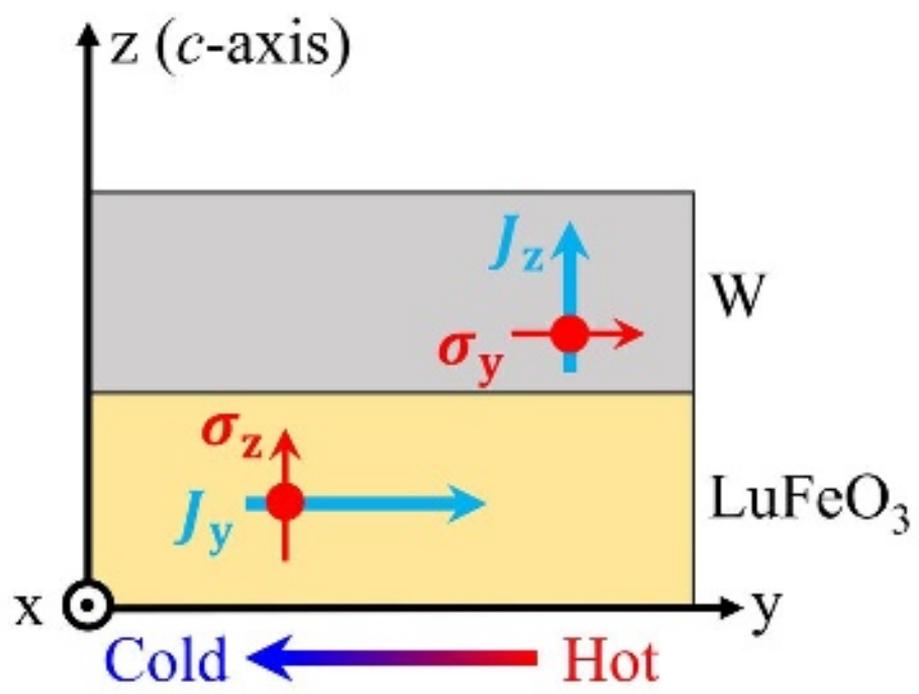

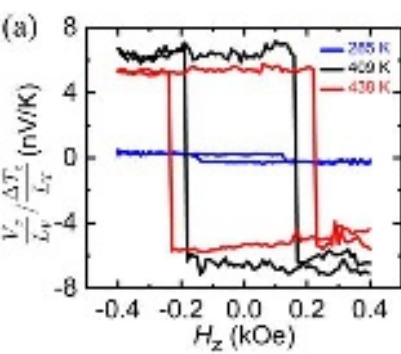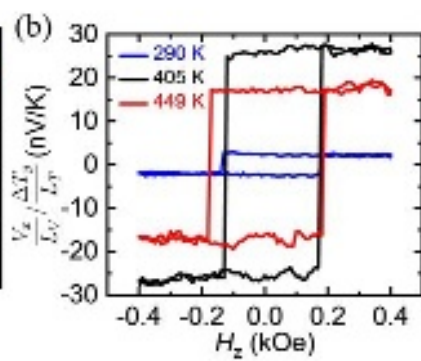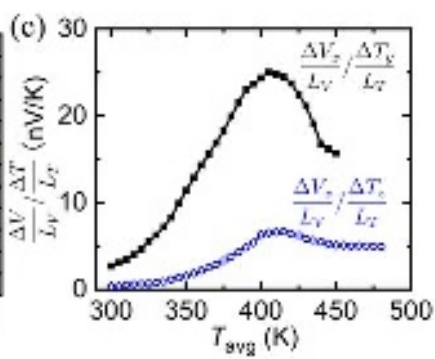